%Paper: chao-dyn/9407017
%From: Jean-Bruno Erismann <jb@cptsu5.univ-mrs.fr>
%Date: Mon, 1 Aug 94 09:52:08 +0200

\magnification=1200
\catcode `@=11

\def\n{\noindent}
\def\m{\medskip}
\def\b{\bigskip}

\font\ten=cmbx10 at 13pt
\font\twelve=cmbx10
\font\eight=cmr8

\font\tencm=cmbxti10
\font\tenll=lasy10
\newfam\llfam
\textfont\llfam=\tenll

\def\max{\mathop{max}}
\def\build#1_#2^#3{\mathrel{\mathop{\kern 0pt#1}\limits_{#2}^{#3}}}

\baselineskip 15pt

{
\centerline{\twelve Centre de Physique Th\'eorique - CNRS - Luminy,
Case 907}
\centerline{\twelve F--13288 Marseille Cedex 9 - France }
\centerline{\bf Unit\'e Propre de Recherche 7061}

\vskip 2truecm

\centerline{\ten STABILITY OF WAVELENGTHS}
\centerline{\ten AND SPATIOTEMPORAL INTERMITTENCY}
\centerline{\ten IN COUPLED MAP LATTICES}

\bigskip

\centerline{
{\bf Andr\'e LAMBERT}\footnote{$^{\star}$}{\eight and
Universit\'e Aix-Marseille II} {\bf and Ricardo LIMA}
}

\vskip 3,5truecm

\centerline{\bf Abstract}

\medskip

In relation to spatiotemporal intermittency, as it
can be observed in coupled map lattices, we study the stability of
different wavelengths in competition.

\noindent Introducing a two dimensional map, we compare its dynamics
with the one of the whole lattice. We conclude a good agreement
between the two.

\noindent The reduced model also allows to introduce an order
parameter which combines the diffusion parameter and the spatial
wavelength under consideration.

\vskip 3,5truecm

\noindent December 1992

\noindent CPT-92/P.2832

\bigskip

\noindent anonymous ftp or gopher : cpt.univ-mrs.fr

\footline={}

\vfill\eject
}

\pageno=1

\baselineskip 18pt

\n{\bf 1. INTRODUCTION}
\m
Coupled Map Lattices (CML) have been proposed as simplest models for
space-time complexity.

Spatially extended dynamical systems are typically present in many
experimental situations.

Examples of such systems can be found in hydrodynamics, plasma
physics,  chemical reactions or biological systems. For those
non-linear systems, CML gives a useful alternative point of view to
the more conventional tool of partial differential equations.

The advantage of CML is based on the use of local maps with well
controlled dynamical properties, that are synchronously updated and
with suitably chosen local or global coupling [1-4].

Different models of CML have been widely investigated from various
point of  view: existence and properties of an invariant measure for
infinite lattices at low coupling [5], stability and universal
properties of homogeneous states [3], [7], [8], [9], study of the
dynamical regime by means of a bi-orthogonal decomposition [10],
[11], [17] and, probably the most important issue in view of the
physical applications as they appear, for instance in
Rayleigh-B\'enard convection [12], the understanding of
spatiotemporal intermittency [1-3] and [13-16]. The emergence of
spatial coherent  structures from a local chaotic dynamics in CML
was discussed in [6]. On the other hand, the coexistence of
different domains in the lattice with coherent dynamical behaviour,
separated by boundaries (kinks), was one of the first and
fascinating features observed in these models.

The relation of this situation with spatiotemporal intermittency has been
mentioned by several authors [1], [4], [15] and was named ''natural
wavelengths'' by the later authors.

In a way, this mechanism can be described as follows: the size of
the different  coherent regions is modified by a slow displacement
of the kinks, until a destabilization of the periodic structure
occurs inside them. Then, after a transient regime of variable
duration, a new periodic state is restored in those regions. In fig.
1-a, b and c we present three plots of the dynamics for typical
situations.

There, we can see spatiotemporal representations of a
one-dimensional lattice - see (2.1) for an exact definition of the
model - for different values of the parameters. The stroboscopic
period $\Delta T=512$ or $256$ is chosen according to the time scale
of the modifications observed in the lattice.

As already pointed out by Kaneko [1], [4] and Keller and Farmer
[15], the slow  motion of the kinks is due to the presence, between
the two regions, of a nodal point $x_i$, lying in the neighbourhood
of the fixed point of the logistic map.

In any case, the spatial wavelength is modulated by the motion of
the domain walls until the destabilization of this wavelength occurs
as a threshold phenomenon.

It will be shown later that bifurcations of the temporal period of
the dynamics may also occur.

It is also obvious that, inside each region, the wave number, an
integer, must be constant between two consecutive unstable
situations. Therefore, changing the size of the region, will involve
a continuous change of the spatial wavelength
$\lambda$, of the corresponding configuration.

{}From this point of view it seems natural to study the stability of
these  wavelengths. They act as a bifurcation parameter in this
transition between local order and chaos as was first observed in
[15], and therefore we guess that it may help in the understanding
of spatiotemporal intermittency.

The aim of the present work is to study analyticaly and numericaly
some features of this mechanism. Let us mention the two main
ingredients used in our paper.

First, we propose a two-dimensional mapping depending on two
parameters, whose dynamics (cycles, bifurcations) may describe the
main features of the whole lattice. In a sense that will become
clearer later, we can say that the whole CML dynamics is enslaved by
the one of this simple model.

The first parameter of the reduced map is related to the
nonlinearity of the local map in the CML (here the logistic map)
describing the dynamics at each site of the lattice. The second
parameter contains the diffusive constant governing the interaction
between neighbouring sites together with the space wavelength and,
therefore, is used as a control parameter.

The second main ingredient is used in numerical simulations.
Instead of trying to describe the ''real'' situation as it is shown
in fig.1, we build an ''artificial'' recipe where the whole lattice
is made of only one region and the stability of the space wavelength
is studied by changing the number of sites in the lattice.

This is done by means of an ''adiabatic compression'' described in
section 3  below and such that
the reader can easily imagine by having a look at fig.2.

The paper is organized as follows:

In section 2, after a brief presentation of the models of interest,
we use the Fourier transform of spatial structures to introduce the
reduced model that is described in section 4. Obviously, this step
is reminescent of amplitude equations as they are used in weak
turbulence [18].

In section 5 we present the simplest case for which the reduced
model is exact. It will be come clear that this case corresponds to
the situation described in [6]. A way of using this reduced model
out of the region of stability, by means of a bi-orthogonal
decomposition was proposed in [17].

In section 6 we deal with the stability in presence of a
time-period two dynamics and in section
7 of period four.

Section 8 deals with the case of homogeneous dynamics.

In section 9 we briefly compare the previous results with numerical
computations of Lyapunov exponents and section 10 concludes.
\vfill\eject
\n{\bf 2. THE MODELS AND THEIR FOURIER ANALYSIS.}
\m
In this work we consider a one-dimensional lattice of length $L$
and a nonlinear map $f$ of the unit interval into itself. The state
of the system at time $t$ in configuration  space is given by
$x_i^t,\quad o\leq x_i^t\leq 1,\quad 1\leq i\leq L$. The new state
at time $t+1$ is defined in each site by the following convex linear
combination of the updated values of neighbour sites :
$$x_i^{t+1}=\left(1-\varepsilon\right)f\left(x_i^t\right)+
{\varepsilon\over
2}\left\{f\left(x_{i+1}^t\right)+f\left(x_{i-1}^t\right)\right\}
\eqno(2.1)$$
that we often will write :
$$x_i^{t+1}=f\left(x_i^t\right)+{\varepsilon\over
2}\Delta f(x_i^t)\eqno(2.2)$$
where $\Delta$ is the discrete Laplacian
operator.

Here $\varepsilon$ is the coupling between neighbour sites, $0\leq
\varepsilon\leq 1$ and $f$ is chosen as the logistic map :
$$f(x)=\mu x\left(1-x\right),\quad 0\leq\mu\leq 4\eqno(2.3)$$

We always consider periodic boundary conditions in (2.1), namely
$x_{L+1}=x_1,$\break $x_L=x_0$.

In the following we shall see that most of the results obtained for
the model (2.1) are still valid for a more general one, where the
coupling between sites is given by a diffusion operator. This case
can be considered as more realistic if one has in mind a limiting
process to obtain a continuous configuration space.

For this model, already introduced in [15], we replace formula
(2.1) by the following :
$$x_i^{t+1}=\sum_{j=-N}^N\rho_j\ f\left(x_{i+j}^t\right)\eqno(2.4)$$
where $N$ is the range of the interaction and the $\rho_j$ are real
positive numbers satisfying : $$\sum_j\rho_j=1\eqno(2.5)$$
and
$$\rho_j=\rho_{-j}\eqno(2.6)$$

Clearly the model described by (2.1) is a particular case of this
one.

In the numerical simulations of (2.4) we have chosen a sequence of
linearly decreasing weights :
$$\rho_j=c\left(1-{\vert j\vert\over N+1}\right),\quad 0\leq\vert
j\vert\leq N+1\eqno(2.7)$$
where $c$ is given by the normalization relation (2.5).

Finally the parameter $\mu $ of the logistic map will always verify
$\mu >\mu_c$, where
$\mu_c=3,5699\dots$ is the critical value  above which the map $f$
has a chaotic dynamics.

In some cases we will have to consider values of $\mu $ such that
$\mu_c<\mu<\mu_1=3,677\dots$ for which the dynamics of $f$ is
attracted inside two disjoint intervals $\Delta_1$ and $\Delta_2$
whose extreme points are determined by the four first iterated by
$f$ of the point $x=1/2$.

Then, every point of $\Delta_1$ (resp $\Delta_2$) is mapped by $f$
in a point of $\Delta_2$ (resp
$\Delta_1$).

For $\mu>\mu_1$ this two intervals overlaps.

Let's now concentrate on the model (2.1).

Since we are interested in the study of space periodic
configurations, it is natural to introduce their discrete Fourier
transforms.

In absence of convective terms, we can restrict ourselves to even
configurations, bringing to the following representation :
$$x_i^t=\sum_{k=0}^{L/2}a_k^t\cos\left({2\pi k i\over L}\right)
\eqno(2.8)$$
{}from where immediatly follows :
$$\Delta x_i^t=\sum_{k=0}^{L/2}2\ a_k^t\left\{\cos\left({2\pi k\over
L}\right)-1\right\}\cos\left({2\pi k i\over L}\right)\eqno(2.9)$$

Finally, using the explicit form of the logistic map, we find the
equations which give the dynamics of the Fourier coefficients.

For the saving of notations we write $a_k=a_k^t$ and $\overline
a_k=a_k^{t+1}$.
\eject
Therefore we get :

\hbox to 16,5cm{\vrule\narrower{\parindent=0,8cm}\vbox{
$$\eqalign{
&\overline a_0=f\left(a_0\right)-{\mu\over 2}\left(\sum_{i=1}^{L/2}
a_i^2+a_{L/2}^2
\right)\hfill\cr
&\overline a_{L/2}=\left(1-2\varepsilon\right)\mu\left\{a_{L/2}
\left(1-2 a_0\right)-{1\over 2}\sum_{k=1}^{L/2-1}a_k\
a_{L/2-k}\right\}\hfill\cr}$$

\n and for $q=1,2,\dots,L/2-1$ :
$$\eqalignno{
\overline
a_q=&\mu\alpha\left(q,\varepsilon\right)\Biggl\{a_q\left(1-2a_0
\right)- {1\over 2}\sum_{k=1}^{q-1}a_k\ a_{q-k}-\hfill\cr
&-{1\over 2}\sum_{k=0}^q a_{k+L/2-q}\
a_{L/2-k}-\sum_{k=1}^{L/2-q}a_{k+q}\
a_k\Biggr\}\hfill&(2.10)\cr}$$}}

\n where
$$\alpha\left(q,\varepsilon\right)=1-2\ \varepsilon\sin^2\left(
{\pi q\over L}\right)\eqno(2.11)$$
We denote by $\cal F$ this $\left(L/2+1\right)$ dimensional map.
Notice that,  for each $q$, the non linear part of $\cal F$ is a
convolution product which is proportional to the sum of the $a_i\
a_j$ where the indices
$\left(i,j\right)$ belong to the boundary of a rectangle in the
$L/2\times L/2$ square lattice of Fourier modes defined by the
corners $\left(0,q\right),\
\left(q,0\right),\ \left(L/2,\ L/2-q\right),\ \left(L/2-q,\
L/2\right)$. This rectangle is reduced to one of the diagonals for
each of the exceptional terms
$\overline a_0$ and $\overline a_{L/2}$.

It is worth to notice that the coupling parameter $\varepsilon$
enters in $\cal F$ only by means of (2.11). Therefore, the relation
(2.11) has the meaning of a rescaling between $\varepsilon$ and the
set of wavenumbers $q$.

In order to motivate the introduction later on of a reduced model
we make now a remark about the relation of equations (2.10) with the
corresponding dynamics in physical space.

The first equation of (2.10) describes the dynamics of the $x_i$
spatial mean value since
$$a_0=<x>={1\over L}\sum_{i=1}^L x_i\eqno(2.12)$$

On the other hand, writing
$$x_i=<x>+\delta_i\eqno(2.13)$$
we can rewrite the first equation in (2.10) as well as the dynamics
of the variances $\delta_i$. Thanks to the chosen boundary
conditions we may have
$<\Delta f(x)>=0$ and therefore we get
$$\eqalignno{\overline{<x>}&=f\left(<x>\right)-\mu<\delta^2>\cr
\overline{\delta}_i &=\mu\left(1-2<x>\right)\left(1+{\varepsilon
\over 2}\Delta\right)\delta_i-\mu\left(1+{\varepsilon\over
2}\Delta\right)\left(\delta_i^2-<\delta^2>\right)&(2.14)}$$
where, again $u,\ \overline u$ stand for the values of $u$ at time
$t$ and $t+1$.

Coming back to (2.10) we also remark that, if at time $t_0$ we
start with a configuration for which the only non vanishing
coefficients are $a_0$ and
$a_{q_0}\left(q_0\le L/4\right)$ then at time $t_0+1$ the only new
created harmonics corresponds to $a_{2q_0}$. Therefore if $q_0$ and
$\varepsilon$ are such that
$$\alpha\left(2q_0,\ \varepsilon\right)=0\eqno(2.15)$$
the only non vanishing amplitude for any time remains $a_{q_0}$.

In other words, in that situation,
$$x_i^t=a_0^t+a_{q_0}^t\cos{2\pi q_0 i\over L}\eqno(2.16)$$
is an exact solution. This is the case, for instance for
$$\lambda_0={L\over q_0}=6\quad\hbox{and}\quad\varepsilon=2/3
\eqno(2.17)$$

We shall see that such a selection of a spatial wavelength by means
of a non linear coupling between the spatial mean value and the
amplitude of the oscillation is observed in much more general
situations than the ones strictly obeying (2.15).

In order to get an insight into this phenomenon it is useful to
look for the behaviour of $\alpha(q,\ \varepsilon)$ as a function of
$q$, for a given $\varepsilon$. As an example this function is
plotted in fig. 3 for
$\varepsilon=0,667$ and $L=100$. The behaviour of $\alpha$ near to
the point where $\alpha=0$ indicates we may expect the damping of
all a set of harmonics.

We can consider the case $\lambda=2$ $\left(q_0=L/2\right)$ to be
very special, since then the system evolves according to only the
first two equations in (2.10) where $a_{L/2}=a_{q_0}$ is coupled
only to the mean value $a_0$. No condition as (2.15) is required for
the existence of an exact solution given by (2.16). We will come
back to this case in section 5.

To end up this section we briefly indicate how the previous results
should be modified if, instead of the simplest model (2.1) we take
the more general diffusive model (2.4).

The dynamics of Fourier coefficients obeys the same equations as
(2.10) where we only need to replace $\alpha\left(q,\
\varepsilon\right)$ by :
$$\alpha(q,\
\{\rho\})=1-4\sum_{j=1}^N\rho_j\sin^2\left({\pi q j\over
L}\right)\eqno(2.18)$$
where $N$ is the range of the diffusion operator. Let us notice
that (2.18) is a good candidate for a limiting process to a
continuous space.

Then, the equations for the dynamics of the mean value and
variances, similar to (2.14), are the following :
\vskip -3mm$$\left\vert\normalbaselineskip=25pt\matrix{
&\overline{<x>}=f(<x>)-\mu<\delta^2>\hfill\cr
&\overline{\delta_i}=\mu(1-2<x>){\cal D}(\delta_i)-\mu{\cal
D}(\delta^2_i-<\delta^2>)\cr}\right.\eqno(2.19)$$
\vskip -1mm
\n where the diffusion operator ${\cal D}$ is defined as

$${\cal D}(x_i)=\sum^N_{j=-N}\rho_jf(x_{j+i})\eqno(2.20)$$
and we used the normalization (2.5) and symmetry (2.6) for the
$\rho_j$ as well as the periodic boundary conditions.

Fig. 4 shows a plot of the function $\alpha(q,\{\rho\})$ where the
special form of the $\rho_j$ given by (2.7) is considered. Compared
with Fig. 3, the faster decay of $\alpha$ as a function of the wave
number $q$ indicates a selection of larger wavelengths in this case
and this is what we can observe numerically.

\vfill\eject

\n{\bf 3. ADIABATIC COMPRESSION : A NUMERICAL RECIPE}

\m

As it was alredy mentioned in the introduction about the "natural"
model, (i.e. the one build on a lattice of constant size $L$), the
spatial wavelengths which are present during the dynamics are
modulated by the motion of the their respective domain walls. Our
aim is to study the instabilities which give rise to these
spatiotemporal oscillations as the source of spatiotemporal
intermittency.

In order to compare this situation in the lattice with the
behaviour of the simplified model presented in Section 4 we propose
to isolate that phenomenon by means of a numerical experiment. We
call it adiabatic compression of the lattice that we will describe
now.

When the lattice of size $L$ reaches a periodic state (eventually
after a transient regime) we carry out the following transformation
:

\item{(i)} $L$ becomes $L'=L-1$

\item{(ii)} $x_i$ becomes $x_i'=x_i+(i-1)(x_{i+1}-x_i)/(L'-1)$

\item{(iii)} the periodic boundary conditions are restored for the
new lattice.
\hfill(3.1)

Notice that this transformation preserves the wavenumber but may
change the corresponding wavelength.

After relaxation we may observe if the new wavelength of spatial
oscillations is stable or not.

We repeat the operation until we notice a transition to a
disordered state. Then the critical value $\lambda_{cr}$ is given by
$\lambda_{cr}=L_1/q$ where $q$ is the wavenumber and $L_1$ the
actual length of the lattice.

After this transition the lattice will relax (with constant length
$L_1$) to a new periodic state with new wavenumber $q_1$ and,
therefore, a wavelength
$\lambda_1=L_1/q_1$. The same process may be repeated now.

Coming back to fig. 2.a we can see there, two adiabatic
compressions followed by the correspondent instability transitions.
\eject
In this example the model is as (2.1) with $\mu=3.63$ and
$\varepsilon=0.667$. The initial number of sites in the lattice is
$L=85$ and the wavenumber is
$q=10$ $\left(\lambda=8.5\right)$. After adiabatic compression the
first destabilization occurs for $L_1=82$ and then
$\lambda_{cr}=8.2$.

After relaxation the new wavenumber is $q_1=7$ and therefore
$\lambda=82/7=11.7$.

After a new adiabatic compression the system undergoes another
destabilization for $L_2=57$ and therefore $\lambda'_{cr}=57/7=8.1$
which now relaxes to a new wavenumber $q_2=5$ which corresponds to
$\lambda'=11.4$.

Fig. 2.b gives more details on the first described adiabatic
compression, since we use now a stroboscopic period $\Delta T=32$
instead of the $\Delta T=256$ used in fig. 2.a. We can see how the
initial perturbation propagates along the lattice and then relaxes
to the new laminar state with $q_1=7$.

One may observe that both destabilizations are obtained for close
values of
\break $\lambda(\lambda_{cr}=8.2$ and
$\lambda'_{cr}=8.1)$ as well as are close the two values of the
spatial wavelengths selected after relaxation $\left(\lambda=11.7\
\hbox{and}\
\lambda'=11.4\right)$.

Notice that $\left(\lambda_{cr}-\lambda'_{cr}\right)$ as well as
$\left(\lambda-\lambda'\right)$ are as small as they are allowed to
by the rational approximation determined by the maximum size of the
lattice in this case and the correspondent integer wavenumbers. So
we may wonder if by taking a much larger lattice at the begining a
better agreement of these values may be found. This is effectively
true, but the price to pay is an extremely larger relaxation time
(transients) and so even for a small gain of precision, we refrain
to pursue in this direction.

Instead we may notice the time period of the dynamics represented
in fig. 2 also undergoes a bifurcation above $\lambda=11$, a fact
that we will adress later on. The reason why this is not observed in
fig. 2 is simply because there, the stroboscopic periods are always
multiples of the time periods of the dynamics.

We will use several times the adiabatic compression to test the
predictions of the simplified model described in the following
section, model where $\lambda$ is taken as the varying control
parameter.
\vfill\eject

\b
\n{\bf 4. A REDUCED MODEL}
\m
We define the following two-dimensional dynamical system :
\vskip -3mm$$\left\vert\normalbaselineskip=30pt
\matrix{
X_{t+1}&=&f\left(X_t\right)-\displaystyle{{\mu\over 2}}Y_t^2
\hfill\cr
Y_{t+1}&=&\alpha\left(\lambda,\varepsilon\right)f'\left(X_t\right)
Y_t\cr}\right.
\eqno(4.1)$$
\vskip -1mm
where $\alpha\left(\lambda,\varepsilon\right)=1-2\varepsilon\sin^2
\left(\pi /\lambda\right)$ and $f$ is defined as the logistic map
with parameter $\mu$ as before.
Let's denote $F$ the map defined by (4.1).
Remark that now $\alpha$ is considered as a function of continuous
parameters
$\lambda$ and $\varepsilon$.

The definition (4.1) comes from a matching of equations (2.10)
describing the dynamics of the Fourier modes and the corresponding
formulation (2.14) in terms of mean and variances.

It can also be deduced from (2.10) by means of the first terms of
an expansion for small amplitudes compared with the mean value, in
which case $X_t$ represents the space mean value of the
configuration at time $t$ and $Y_t$ the global amplitude of the
variance at the same time.

Even if the real meaning of $Y_t$ is to be taken with caution, we
got a good agreement of the values of $Y_t$ according the map (4.1)
and the variance of the configuration at same time as it can be
computed form direct numerical simulation with the whole lattice.

The only exception is for $\lambda=2$ where the factor $\mu/2$ in
the first equation of (4.1) must be replaced by $\mu$ if we want to
maintain this interpretation for $Y$. This is due to the periodic
boundary conditions. Instead of changing the form of the map $F$ for
this special case $\lambda=2$, we prefer to keep it in the same form
as (4.1) since this simply corresponds to a rescaling
$Y'=Y/\sqrt{2}$ which obviously do not change the properties of
(4.1).

Any way, we take (4.1) as a starting point and we will show how it
behaves on the parameter space. Different bifurcations and dynamical
regimes of (4.1) will be interpreted and then compared to the
corresponding ones for the whole lattice.

Notice the map $F$ depends on three parameters : $\mu,\
\varepsilon$ and
$\lambda$, but the later two enter only in (4.1) by means of the
function
$\alpha(\lambda,\varepsilon)$. Therefore $F$ really depends only on
two parameters : $\mu$ and $\alpha$.

We shall compute different cycles of $F$ (for different temporal
periods $T$) and determine their stability domains in parameter
space.

When expressed as a function of $\varepsilon$ the result will be
compared to numerical simulations showing bifurcation diagrams of
the whole CML as a function of $\varepsilon$.

On the other hand, as already written, when the results are
expressed as a function of $\lambda$, they will be confirmed by
applying to the CML the adiabatic compression defined in Section 3.

Before going in the more detailed study of the various dynamical
regimes of the map $F$, we notice a special feature of it that helps
to understand some of the results described below.

The second equation in (4.1) only fixes the ratio $Y_{t+1}/Y_t$ and
terefore is invariant by scaling of the variable $Y$ as far as the
condition $0\leq Y\leq1$ is fulfilled. This allows the translation
described by the first equation to stabilize the logistic map
governing the dynamics of $X$.

Then, the coupling between the two variables is closed because, in
order that $Y_t$ stays in a bounded domain, this stable orbit for
$X$ should be such that $\alpha f'(X_t)$ oscillates around 1 (the
case where it is always, in modulus, strictly less than 1 is
addressed in Section 8).

Therefore, at the end we see that the dynamics is governed by a
balance of the size of the oscillations of the mean variable $X$ and
the one of the variance amplitude that may stabilize each other.
\vfill\eject

\b
\n{\bf 5. THE SIMPLEST CASE.}
\m
We begin our analysis with the simplest dynamics of map (4.1),
namely fixed points $\left(T=1\right)$.

The coordinates of such a fixed point obey the following system of
equations :
\vskip -3mm$$\left\{\normalbaselineskip=30pt\matrix{
X&=&f(X)-\displaystyle{{\mu\over 2}}Y^2\hfill\cr
Y&=&\alpha f'(X).\ Y\hfill\cr}\right.\eqno(5.1)$$
\vskip -1mm
{}From the second equation of (5.1) we can see that an obvious fixed
point holds with $Y=0$. But then we discover from the first equation
that it corresponds to the (unstable) fixed point of the logistic
map, $X=f(X)$. According to the previous section, this should be an
homogeneous state (with vanishing variance) and it is easy to show
it is not stable. In section 8, we will come back to this issue.

Non obvious solutions $\left(Y\ne 0\right)$ are easily computed as :
\vskip -1mm$$\left\{\normalbaselineskip=35pt\matrix{
X &=&\displaystyle{{1\over 2}\left(1-{1\over\alpha\mu}\right)}
\hfill\cr
Y &=&\displaystyle{\left\{{1\over
2\mu}\left(\mu-2+{1\over\alpha}\right)
\left(1-{1\over\alpha\mu}\right)\right\}^{1/2}}\hfill\cr}
\right.\eqno(5.2)$$
\m
Here $\alpha=\alpha\left(\lambda,\ \varepsilon\right)=1-2
\varepsilon\sin^2
\left({\pi\over\lambda}\right)$. Notice we are only interested in
fixed points for which $0\le X\le 1$ and $Y^2\le 1$. These
conditions fix the boundaries, in parameter space, for the existence
of solutions with $T=1$.

The conditions for the stability of such fixed points are found by
computing the eigenvalues of the tangent map of (4.1) evaluated on
the solutions given by (5.2) :
\vskip -3mm$$Q_1=\left\vert\normalbaselineskip=25pt\matrix{
1/\alpha\hfill&-\mu Y\cr
-2\alpha\mu Y&1\hfill\cr}\right\vert\eqno(5.3)$$
\vskip -1mm
As it can be easily shown, for the values of $\mu$ that we are
interested in and unless $\lambda=2$ or $\lambda=3$, the stability
of these fixed points will imply $\varepsilon>1$.
\eject
We have numerically confirmed the predictions also in this case,
but we refrain to present them here since they go beyond the frame
of the present work, see [19].

We then focus on the simplest case, $\lambda=2$, in order to show
we can recover the results shown in [6].

First we observe that :
$$\alpha\left(2,\ \varepsilon\right)=1-2\varepsilon\eqno(5.4)$$
and by changing variables :
\vskip -3mm
$$\left\{\normalbaselineskip=30pt\matrix{
A &=& X+\displaystyle{{Y\over
\sqrt{2}}}\hfill\cr
B &=& X-\displaystyle{{Y\over\sqrt{2}}}\hfill\cr}
\right.\eqno(5.5)$$
\vskip -1mm
\n the equations of the fixed points of (4.1) read :
\vskip -3mm$$\left\{\normalbaselineskip=25pt\matrix{A
&=&\left(1-\varepsilon\right)f(A)+\varepsilon f(B)\hfill\cr
B &=&\left(1-\varepsilon\right)f(B)+\varepsilon
f(A)\hfill\cr}\right.\eqno(5.6)$$
\vskip -1mm
\n so that (5.1) reduces to a well known form. $A$ and $B$ are the
values of the configuration for even and odd sites on the lattice
and, therefore, the existence of a solution for (5.6) is equivalent
to the one of the given type for the whole lattice.

In other words, as already mentioned in section 3, the reduced map
(4.1) is exact when $\lambda=2$.

{}From (5.6) we easily find that $A$ and $B$ are the roots of the
equation :
$$u^2-S u+P=0\eqno(5.7)$$
where
$$S=A+B=1-{1\over\mu\left(1-2\varepsilon\right)}\eqno(5.8)$$
and
$$P=A B={-\varepsilon\over\mu\left(1-2\varepsilon\right)}
\left(1-{1\over\mu\left(1-2\varepsilon\right)}\right)\eqno(5.9)$$

\n which obviously agree with (5.1) where $\alpha$ is taken as in
(5.4).
\eject
On the other hand, the stability of such fixed points is given by
the condition
$$\vert\nu_{\pm}\vert<1\eqno(5.10)$$
where $\nu_{\pm}$ and the two roots of the usual equation deduced
{}from the jacobian of (5.6) \break namely :
$$\nu^2-\hbox{\tencm T}\nu+\hbox{\tencm D}=0\eqno(5.11)$$
where :
$$\left\vert\normalbaselineskip=25pt\matrix{\hbox{\tencm T}
&=&\left(1-\varepsilon\right)\left\{f'(A)+f'(B)\right\}\hfill\cr
\hbox{\tencm D}&=&\left(1-2\varepsilon\right)\left\{f'(A).\
f'(B)\right\}\hfill\cr}\right.\eqno(5.12)$$
with $f'(x)=\mu\left(1-2x\right)$.

In fig. 5 we show the diagram of bifurcations in parameter space
$\left(\mu,\ \varepsilon\right)$, where the region corresponding to
stable fixed points is filled in grey.

We also mention it is possible to show that the stability of the
solution of (5.6) implies the stability of the corresponding
solution for the whole lattice (the proof is given in the Appendix
at the end of the paper).

For that purpose, we notice that the eigenvalue equations for the
tangent map of the CML with a lattice of even length, calculated at
the fixed point defined above may be reduced, by symmetry, to a set
of $L/2$ eigenvalue problems for 2 x 2 matrices. Finally, inspecting
the different cases, we found that, inside the stability domain of
the reduced system all eigenvalues of the CML tangent  map are, in
modulus, strictly less than 1. Furthermore, the two bifurcations of
(5.6) are of different nature. For the lower end point of the
stability interval the eigenvalues are real and they are complex for
the upper end point.

Let us add that, in principle, it is possible to follow the same
lines of equations (5.6) for solutions of the CML with $T=1$ and
$\lambda>2$. This leads to a set of $\lambda$-equations replacing
(5.6) that was studied in [19], but the system became rapidly
cumbersome and it can only be studied by numerical computation.

The advantage of our reduced model (4.1) comes from the fact that
we always have a two-dimensional system, no matter of the wavelength
we consider.
\vfill\eject

\b
\n{\bf 6. TIME-PERIOD TWO DYNAMICS}
\m
We come back again to the reduced model $F$ defined in (4.1) to
analyse the case of solutions with time period 2.

Numerical simulation shows the existence of period 2 cycles
satisfying the following symmetry relation :
\vskip -3mm$$\left\vert\normalbaselineskip=25pt\matrix{X_{t+1}&=&
\hfill X_t\cr
Y_{t+1}&=& -Y_t\hfill\cr}\right.\eqno(6.1)$$
\vskip -1mm
\n No other cycle of period 2 was found for $F$.

\n The coordinates $X,\ Y$ of such cycles obey then the equations :
\vskip -3mm$$\left\vert\normalbaselineskip=25pt\matrix{
X &=& f(X)-\displaystyle{{\mu\over
2}}Y^2\hfill\cr
Y &=& -\alpha f'(X).\ Y\hfill\cr}\right.\eqno(6.2)$$
\vskip -1mm
\n where, again, the dependance of
$\alpha=1-2\varepsilon\sin^2\left({\pi\over\lambda}\right)$ in
$\lambda$ and
$\varepsilon$ has been omitted.

The solution of (6.2) is immediately found :
\vskip -3mm$$\left\vert\normalbaselineskip=30pt\matrix{X &=&
\displaystyle{{1\over 2}\left(1+{1\over\alpha\mu}\right)}\hfill\cr
Y &=&\displaystyle{\left\{{1\over
2\mu}\left(\mu-2-{1\over\alpha}\right)\left(1+{1\over\alpha\mu}
\right)\right\} ^{1/2}}\hfill\cr}\right.\eqno(6.3)$$
\vskip -1mm
The corresponding tangent map is, according to (4.1) given by :
\vskip -3mm$$D F\left(X,\ Y\right)=\left[\normalbaselineskip=25pt
\matrix{ f'(X)\hfill&-\mu Y\hfill\cr
\alpha f"(X).\ Y&\hfill\alpha f'(X)\cr}\right]\eqno(6.4)$$
\vskip -1mm
\n and therefore, for $Q_T=Q_2=D F\left(X,\ -Y\right).\ D F
\left(X,\ Y\right)$, one has :
\vskip -3mm$$Q_2=\left[\normalbaselineskip=30pt\matrix{
\displaystyle{{1\over\alpha^2}-2\alpha
\mu^2 Y^2}&-\mu\displaystyle{\left(1-{1\over\alpha}\right)Y}\cr
-2\mu\left(1-\alpha\right)Y\hfill&1-2\alpha\mu^2 Y^2\cr}
\right]\eqno(6.5)$$
\vskip -1mm
\n and then the determinant and trace of $Q_T$ are :
\eject
$$\left\{\normalbaselineskip=30pt\matrix{\det Q_2
&=&\displaystyle{\left({1\over\alpha}-2\alpha\mu^2 Y^2\right)^2}
\hfill\cr
T_r\ Q_2 &=&\displaystyle{1+{1\over\alpha^2}-4\alpha\mu^2
Y^2}\hfill\cr}\right.\eqno(6.6)$$
\vskip -1mm
\n from which we can compute the eigenvalues
$\nu_{\pm}$ as roots of the corresponding usual equation.

As an example of the result accuracy comparing to the behaviour of
the CML, we first show a case where $\mu=3.8$ and $\lambda=6$ are
keeped fix and
$\varepsilon$ varies. For this case fig. 6-a shows the largest
value of the modulus of the eigenvalues of $Q_2$ as a function of
$\varepsilon$. This function is equal to $1$ for
$\varepsilon_1=0.39\dots$ in the domain where the eigenvalues are
complex and for $\varepsilon_2=0.695\dots$ where they are real.

Fig. 6-b shows the bifurcation diagram of a CML of 60 sites for
which the two corresponding values of bifurcation are
$\varepsilon'_1=0.4\dots$ and
$\varepsilon'_2=0.684$. As explained in the introduction larger
lattice gives an even better agreement.

Fig. 6-c gives, for the same interval in $\varepsilon$, the
bifurcation diagram of the second coordinate $-\ Y\ -$ of the
two-dimensional map $F$, where we can see the relation between $F$
and the dynamics of the CML is deeper than the only coincidence of
the two bifurcation points. This relation will be treated elsewhere
using the method described in [17].

As a second test we take $\varepsilon=0.667$ to be fixed as well as
$\mu=3.8$, the same value as before and taking now $\lambda$ as
order parameter.

The largest value of the modulus of the eigenvalues of $Q_2$ is now
plotted as a function of $\lambda$ in fig. 7.

We found $\lambda_1=5.96$ and $\lambda_2=7.98$ for the
corresponding values of bifurcation. Notice that, due to the form of
the function
$\alpha\left(\lambda,\ \varepsilon\right),\ \lambda_1$ corresponds
to
$\varepsilon_2$ and $\lambda_2$ to $\varepsilon_1$.

Following the discussion of section 3 we compare now these values
with the actual CML bifurcation diagram by using an adiabatic
compression.

We begin with a snapshot of a lattice which length is $L=63$ and a
wavenumber
$q=9$ $\left(\lambda=7\right)$ as shown in fig. 8-a. Fig. 8-b shows
the lattice after adiabatic compression until $L=55$, where we can
see the relaxed state with the same wavenumber is still stable. The
\eject
destabilization occurs for
$L=54\ \left(\lambda'_1=6\right)$ as it is seen in fig. 8-c. Then
the system, relaxes always with $L=54$, to a new stable state as it
is shown in fig. 8-d after a transient regime of about 4000
iterations. Fig. 8-e shows a superposition of two such states at
time $t$ and $t+2$. Notice that in fig. 8-d and 8-e we are concerned
with states of temporal period 4 and a more complex spatial
behaviour which is reminescent of the Eckhaus instability in
Rayleigh-B\'enard convection [18].

Finally, with reference to fig. 8-f and 8-g, applying again the
adiabatic compression a state with $T=2$ is restored for
$L=48\ \left(\lambda'_2=8\right)$ showing again a good agreement
with the prediction of the reduced map $F$.

The regions of parameter space where the map $F$ has stable periodic
$\left(T=2\right)$ dynamics are shown in fig. 9 in the plane
$\left(\mu,\ \varepsilon\right)$ for different values of
$\lambda\left(\lambda=2,4,6,8\right)$ and in fig. 10 in the plane
$\left(\lambda,\ \varepsilon\right)$ for $\mu=3.8$.

We have made a large number of numerical simulations for the CML
and found they were also in good agreement with the predictions of
$F$ as they are shown in these two pictures.

We end up this section by adding that, as already noticed, changing
$\varepsilon$ in $\left(1-\varepsilon\right)$ in equations (5.4)
transforms a fixed point in a cycle of period 2. We notice that this
is a special case of a more general transformation that takes
equation (5.1) of a fixed point of $F$ in equation (6.2) of a cycle
of period 2, by changing $\alpha$ in $-\alpha$.
\vfill\eject

\b
\n{\bf 7. TIME-PERIOD FOUR DYNAMICS}
\m
Exactly in the same way of section 6 we can study the solutions
with period 4 of $F$ and then to compare the dynamics of $F$ with
CML one.

Again we observed that the cycles have the symmetry $Y\to -Y$ of
the map $F$, so we only need to describe the cycle four positive
numbers $X_1,\ X_2,\ Y_1,\ Y_2$, cycle that is runned in the
following way :
$$\left(X_1,\ Y_1\right)\to\left(X_2,\ -Y_2\right)\to\left(X_1,\
-Y_1\right)\to
\left(X_2,\ Y_2\right)\to\left(X_1,\ Y_1\right)\eqno(7.1)$$
This combines the permutation $X_1\to X_2$ between $\Delta_1$ and
$\Delta_2$ for the logistic map described in section 1 and the flip
of period $T=2$ studied before.

Therefore a cycle $T=4$ is solution of
$$F^2\left(X,\ Y\right)=F\left(X,\ -Y\right)\eqno(7.2)$$
or, in other words, solution of the system :
\vskip -3mm$$\left\{\normalbaselineskip=30pt
\matrix{\displaystyle{X=f\left(f(X)-{\mu\over 2}Y^2\right)-{\mu
\over 2}\left(\alpha.\ f'(X).\ Y\right)^2}\hfill\cr
\displaystyle{\alpha^2.\ f'\left(f(X)-{\mu\over
2}Y^2\right).\ f'(X)=-1}\hfill\cr}\right.\eqno(7.3)$$
\vskip -1mm
This leads, after straightforward calculations, to the following
equation, expressed in the variable $Z=\alpha.\ f'(X)$ :
$$Z^2\left(Z^4+\alpha^2\mu\left(2-\mu\right)\left(Z^2-1\right)+4
\alpha Z\right)-1=0\eqno(7.4)$$
Denoting $Z_1,\ Z_2$ the two real roots of (7.4) we finally get :
\vskip -1mm$$\left\{\normalbaselineskip=30pt\matrix{X_{1,2}&=
&\displaystyle{{1\over
2}\left(1-{Z_{1,2}\over\alpha\mu}\right)}\hfill\cr
Y_{1,2}&=&\displaystyle{\left\{{2\over\mu}
f'\left(X_{1,2}\right)-{1\over\alpha^2\mu^2
f'\left(X_{1,2}\right)}-{1\over\mu}\right\}^{1/2}}\cr}
\right.\eqno(7.5)$$
\m
We observe that, taking $Z=\pm 1$ we recover the equivalent
expressions for
$T=1$ and $T=2$, which is naturally related with the symmetry of
the cycles and the definition of $Z$ as the ratio of two successive
values of $Y$.
\eject
The stability of such cycles is studied as before.

As in the previous section we only give two examples of application
since all the other cases we have performed show the same accuracy.

In both cases we take $\mu=3.63$, a value for which the attractor
of $f$ lies inside two disjoint intervals $\Delta_1,\ \Delta_2$ as
mentioned before.

First we take $\lambda=8$ fixed and $\varepsilon$ as a parameter.
The stability interval of the cycle of $F$, as it is shown in fig.
11-a, has limit points $\varepsilon_1=0.42$ and $\varepsilon=0.695$
whereas the values obtained from the bifurcation diagram of the CML,
shown in fig. 11-b, gave
$\varepsilon'_1=0.41$ and $\varepsilon'_2=0.69$.

Then, as in section 6, we fix $\varepsilon=0.667$ and use $\lambda$
as a control parameter. From the stability of $F$ we get stability
between
$\lambda_1=7.85$ and $\lambda_2=10.15$ as it is shown in fig. 12,
whereas the adiabatic compression shown in fig. 13 give, for the
critical values of the wavelengths, $\lambda'_1=65/8=8.1$ and
$\lambda'_2=52/5=10.4$.

\n The same comments as in case $T=2$, can be done for $T=4$.

We also notice that, when the eigenvalues of $Q_T$ cross the unit
circle with complex values, the whole lattice undergoes a period
doubling bifurcation.

Finally we want to add that we performed the corresponding numerical
simulations for the values of the parameter for which the reduced
map $F$ has cycles of higher order and we also find a good agreement
with the corresponding dynamics of the CML.
\vfill\eject

\b
\n{\bf 8. HOMOGENEOUS DYNAMICS}
\m
A simple observation about the dynamics of the reduced map $F$ is
worth to be noticed.

Since
$$\sup_X\vert f'(X)\vert=\mu\eqno(8.1)$$
we see that the condition
$$\vert\alpha\vert.\ \mu<1\eqno(8.2)$$
implies, using the second equation of $F$ that $Y_t\to 0$ when
$t\to\infty$ for all initial conditions $\left(X_0,\ Y_0\right)$.

Therefore, in this case, the asymptotic dynamics is defined as
\vskip -3mm$$\left\{\normalbaselineskip=25pt\matrix{X_{t+1}&=&
f\left(X_t\right)\hfill\cr
Y_t\hfill &=& 0\hfill\cr}\right.\eqno(8.3)$$
\vskip -1mm
According to the given interpretation of these variables, such a
dynamics may cor-\break respond to a dynamics of the CML for which
there is synchronized motion (since $Y=0$).

The corresponding dynamics is obviously given by the one of the
logistic map, according to the first equation.

This regime is already observed in CML, see [20], but suprisingly,
the reduced model again gives quite accurate predictions in this
case. We will address this question elsewhere.

Before ending this section, let us add the following comment :
condition (8.2) is fulfilled if
$${1+\mu\over 2\mu}>\varepsilon\sin^2\left({\pi\over\lambda}
\right)>{\mu-1\over 2\mu}\eqno(8.4)$$
which means that, for all $\varepsilon>0$ there is always a
$\lambda\left(\varepsilon\right)$ such as for
$\lambda>\lambda\left(\varepsilon\right)$ (8.4) may not be true.
But this effect can not be seen if the size of the lattice is not
large enough and, moreover, only for an infinite lattice we should
have no such cutoff. At least in the view of the reduced model, we
see some care need to be taken when dealing with finite size
lattices instead of the infinite limit case.
\vfill\eject

\b
\n{\bf 9. LYAPUNOV EXPONENTS}
\m
Finally, let us present some results obtained by computing maximal
Lyapunov exponents along orbits initialized in the vicinity of the
spatiotemporal periodic states which have been studied above.

The aim is to give a more precise meaning for the terms ``chaotic''
and ``laminar'' used for a phase of a CML, here simply defined by
its positive or negative Lyapunov exponents.

Again we compare the results obtained from the dynamics of the
reduced model
$F$ with the ones coming from $\cal F$ together with the explicit
formulae of his tangent map.

Fig. 14-a and b shows a particular typical case for $F$ and $\cal
F$. Here
$\mu=3.8$ and $\lambda=6$ whereas $\varepsilon$ is taken as control
parameter. This is the situation described in section 6, where $T=2$.

We may observe, again, a precise connection between the two
stability intervals. Moreover, for a smaller interval
$\left(0.33..,\ 0.40..\right)$ we see that the Hopf bifurcation of
the map $F$ corresponds to the period doubling of $\cal F$ that we
have already noticed.

Many other numerical computations we have performed, for different
values of the parameters, confirm this agreement between the two
dynamics.
\vfill\eject

\b
\n{\bf 10. CONCLUDING REMARKS}
\m
In this work we were concerned with the problem of spatiotemporal
intermittency as it can be observed in CML. Directed by the idea
that one of the mechanisms undersetting this phenomenon is a
competition of local configurations with different wavelengths we
have addressed the question of the stability of such wavelengths.

The essential point of our work came from the possibility of
describing many features of this bifurcation diagram as well as of
the corresponding dynamics of the CML by means a two-dimensional
map.

The study of this map allows, in a very simple way, to predict the
existence and stability of large ordered subsets of the lattice.

This approach propounds a particular coupling between the spatial
mean value and the variance of the variables in phase space. It also
suggests a particular form of a control parameter that combines the
diffusion constant and the spatial wavelength of the configuration.

Even if this reduced two-dimensional model gives very accurate
predictions for the CML dynamics we were not able, in the general
case, to achieve a complete proof of the correspondance, as it is
the case for $\lambda=2$.

Further progress will lead to the problem of the study of the full
system
$\cal F$ as it was described in section 2 and would give a more
serious basis to the knowledge of the conditions of applicability of
the two-dimensional map
$F$.

\vfill\eject

\n{\bf APPENDIX :}

\m

We show there the stability of state (5.5) of the whole lattice in
the stability domain of the variable $\varepsilon$ of the
corresponding state of the reduced map (5.1).

Due to the fact that the tangent map $D{\cal F}$ of the whole
lattice evaluated on the state (5.5) commutes with the operator of
cyclic permutations of order two, whose eigenvalues are the roots of
unity : ${\cal W}_k=e^{4\pi ki\over L}$,
$k=1,2,\dots,L/2$ ; the eigenvalues problem for $D{\cal F}$ may be
easily reduced to the one of the set of $L/2$ matrices $M_k$ defined
by :
\vskip -3mm$$M_k=\normalbaselineskip 30pt\left[\matrix{
(1-\varepsilon)f'(A)\hfill&\displaystyle{\varepsilon f'(B)
\left(1+{1\over{\cal W}_k}\right)}\cr
\varepsilon f'(A)(1+{\cal
W}_k)&(1-\varepsilon)f'(B)\hfill\cr}\right]\eqno(A.1)$$
\vskip -1mm
We remark that $M_{L/2}$ is equivalent to the Jacobian of the
reduced map whose trace {\tencm T} and determinant {\tencm D} are
given by (5.12), and the traces
$\hbox{\tencm T}_k$ and determinants $\hbox{\tencm D}_k$ of $M_k$
may be expressed in terms of {\tencm T} and {\tencm D} :
$$\hbox{\tencm T}_k=\hbox{\tencm T}={2(1-\varepsilon)\over
1-2\varepsilon}\ ;\ \hbox{\tencm D}_k=\beta_k\cdot \hbox{\tencm D}
\eqno(A,2)$$
where : $$\beta_k=1+{\varepsilon^2\over
1-2\varepsilon}\cdot sin^2\left({2\pi k\over L}\right)\eqno(A,3)$$

As shown in fig. 5, for $\mu>3.45$, the stability domain of the
reduced map extends from $\varepsilon>0.8$. For such values the
trace {\tencm T} is always negative and $\beta_k$ satisfye for any
$k$ in  $1,\dots,L/2$, the relations :
$$|\beta_k|\leq 1\eqno(A,4)$$
$$-\beta_k\leq\gamma<1\eqno(A,5)$$
where : $\gamma=-\beta_{L/4}=-{(1-\varepsilon)^2\over 1-2
\varepsilon}$.

Now, with the following notations :
$$\eqalign{
&|\nu|=max(|\nu^+|,|\nu^-|),\cr
&|\nu_{\star}|=max(\max_k|\nu^+_k|,\max_k|\nu^-_k|),\cr}$$
\eject
where $\nu^{\pm}_k$ (resp. $\nu^{\pm}$) are the eigenvalues of
matrices
$M_k$ (resp. $M_{L/2}$).

\n and : $\tau=-\hbox{\tencm T},\ \Delta=\tau^2-4 \hbox{\tencm D},\
\Delta_k=\tau^2-4\beta_k\hbox{\tencm D}$

We are going to consider successively the cases :

\n {\bf 1./} $D\leq 0$ and {\bf 2./} $D>0$ with subcases

\n {\bf 2.1/} $\Delta\geq 0$ and {\bf 2.2/} $\Delta<0$, and show in
each of them that, under the hypothesis $|\nu|<1$, one finds :
$|\nu_{\star}|<1$.
\m
\n{\bf 1./} If $\hbox{\tencm D}\leq 0$ then : $|\nu|={\tau
+\sqrt{\tau^2+4|
\hbox{\tencm D}|}\over 2}$ and,
{}from : $\Delta_k=\Delta+4(\beta_k-1)|\hbox{\tencm D}|$ and (A,4) it
follows :
$\Delta_k\leq\Delta$ for any $k=1,2,\dots,L/2$.

Let us denote by $K_1$ (resp. $K_2$) the subset of the indices $k$
where
$\Delta_k\geq0$ (resp. $\Delta_k<0$) and consider separately $K_1$
and $K_2$ :

In the first case : $|\nu^{K_1}_{\star}|=\max\limits_{k\in
K_1}\left({\tau+\sqrt{\Delta_k}\over 2}\right)\leq|\nu|$.

In the second, where the eigenvalues are complex, one has :
$|\nu^{K_2}_{\star}|=\max\limits_{k\in
K_2}\left(\sqrt{\beta_k\hbox{\tencm D}}\right)\leq\sqrt{|
\hbox{\tencm D}|}<|\nu|$.
\m
\n{\bf 2./} If $\hbox{\tencm D}>0$ let us consider separately the
cases of real and complex eigenvalues $\nu$ :

{\bf 2.1/} if $\Delta=\tau^2-4\hbox{\tencm D}\geq 0$, by use of
$\Delta_k=\Delta+4\hbox{\tencm D}(1-\beta_k)$ and (A.4) one finds :
$\Delta_k\geq 0$ and,
consequently :
$|\nu_{\star}|=\max\limits_{k}\left({\tau+\sqrt{\tau^2-4\beta_k
\hbox{\tencm D}}\over 2}\right)$, then, with (A,5) and the relation
$\hbox{\tencm D}<{\tau^2\over 4}$ one obtains for $|\nu_{\star}|$ :

$|\nu_{\star}|\leq\tau\left({1+\sqrt{1+\gamma}\over 2}\right)$ and,
with explicit form of $\tau$ and $\gamma$ as functions of
$\varepsilon$ :
$|\nu_{\star}|\leq{1-\varepsilon\over
2\varepsilon-1}\left(1+{\varepsilon\over\sqrt{2\varepsilon-1}}
\right)<1$.
\m
{\bf 2.2/} for $\Delta<0$ one has $|\nu|=\sqrt{\hbox{\tencm D}}<1$,
we again consider separately the subset $K_3$ and $K_4$ of indices
$k$ for which
$\Delta_k\geq 0$ and $\Delta_k<0$.
\m
\hskip 8truemm{\bf 2.2.1/} if $\Delta_k\geq 0$, using again (A.5) :
$|\nu^{K_3}_{\star}|=\max\limits_{k\in
K_3}\left({\tau+\sqrt{\tau^2-4\beta_k\hbox{\tencm D}}\over
2}\right)\leq{\tau+\sqrt{\tau^2+4\gamma}\over 2}$ and, as a
function of
$\varepsilon$ : $|\nu^{K_3}_{\star}|\leq{1-\varepsilon\over
2\varepsilon-1}(1+\sqrt{2\varepsilon})<1$.
\m
\hskip 8truemm{\bf 2.2.2/} for $k$ such that $\Delta_k<0$ $(k\in
K_4)$ one has :
$|\nu^{K_4}_{\star}|=\max\limits_{k\in
K_4}\left(\sqrt{\beta_k\hbox{\tencm D}}\right)\leq
\sqrt{\hbox{\tencm D}}=|\nu|$  which ends the proof.

\vfill\eject

\n{\bf REFERENCES}
\m

\parindent=1cm
\item{\hbox to\parindent{\enskip [1]}\hfill}K. Kaneko, Prog. Theor.
Phys. {\bf 72} (1984) 480 ; {\bf 74} (1985) 1033.

\item{\hbox to\parindent{\enskip [2]}\hfill}K. Kaneko, Collapse of
Tori and genesis of chaos in dissipative systems. (World Scientific,
Singapore 1896).

\item{\hbox to\parindent{\enskip [3]}\hfill}J.P. Crutchfield and K.
Kaneko, Phenomenology of spatiotemporal chaos in : Directions in
Chaos (World Scientific, Singapore, 1987).

\item{\hbox to\parindent{\enskip [4]}\hfill}K. Kaneko, Physica {\bf
23D} (1986) 436-447.

\item{\hbox to\parindent{\enskip [5]}\hfill}L.A. Bunimovich and
Ya.G. Sinai, Nonlinearity {\bf 1} (1988) 491.

\item{\hbox to\parindent{\enskip [6]}\hfill}L.A. Bunimovich, A.
Lambert and R. Lima, Journal of Statistical Physics {\bf 61} (1990)
253.

\item{\hbox to\parindent{\enskip [7]}\hfill}S.P. Kuznetsov and A.S.
Pikovsky, Physica {\bf 19D} (1986) 384.

\item{\hbox to\parindent{\enskip [8]}\hfill}T. Yamada and H.
Fujisaka, Prog. of Theor. Phys. {\bf 72} (1984) 885.

\item{\hbox to\parindent{\enskip [9]}\hfill}I.S. Aranson, V.S.
Afraimovich and M.I. Rabinovich, Nonlinearity {\bf 3} (1990) 639).

\item{\hbox to\parindent{\enskip [10]}\hfill}N. Aubry, R. Guyonnet
and R. Lima, J. Nonlinear Science, 2 (1992).

\item{\hbox to\parindent{\enskip [11]}\hfill}R. Guyonnet and R.
Lima : "Complex dynamics in Coupled Map Lattices via the
Bi-Orthogonal decomposition". Preprint Marseille 1990.

\item{\hbox to\parindent{\enskip [12]}\hfill}S. Ciliberto and P.
Bigazzi, Phys. Rev. Letters {\bf 60} (1988) 286.

\item{\hbox to\parindent{\enskip [13]}\hfill}H. Chat\'e and P.
Manneville, Physica {\bf 32D} (1988) 409 ; Physica {\bf 37D} (1989)
33.

\item{\hbox to\parindent{\enskip [14]}\hfill}G.L. Oppo and R.
Kapral, Phys. Rev. A {\bf 33} (1986) 4219.

\item{\hbox to\parindent{\enskip [15]}\hfill}J.D. Keeler and J.D.
Farmer, Physica {\bf 23D} (1986) 413.

\item{\hbox to\parindent{\enskip [16]}\hfill}K. Kaneko, Physica
{\bf 37D} (1989) 60.

\item{\hbox to\parindent{\enskip [17]}\hfill}R. Lima, Chaos {\bf 2}
(1992) 315.

\item{\hbox to\parindent{\enskip [18]}\hfill}P. Manneville :
Dissipative Structures and Weak Turbulence (Academic Press).
\eject
\item{\hbox to\parindent{\enskip [19]}\hfill}A.I. Rakhmanov and N.K.
Rakhmanova, On Dynamical Systems with Space Interactions, Preprint
Keldysh Inst. for Applied Math., Moscow, 1990.

\item{\hbox to\parindent{\enskip [20]}\hfill}L. Battiston, L.A.
Bunimovich and R. Lima, Complex Systems {\bf 5}, 415 (1991).
\vfill\eject

\vfill\eject

\n{\bf FIGURES CAPTIONS}

\m
\n{\bf Fig. 1} : Three dimensional views of spatiotemporal
evolution of a CML of 100 sites ($x$ axis) obtained by 100
successive plots ($t$ axis). Half of the lattice has been
initialized in $\Delta_1$ (resp. $\Delta_2$) with
$\varepsilon=0.667$. a) $\mu=3.6$ and the plotting time period is
${\cal T}=512$. b) $\mu=3.62$ and ${\cal T}=512$. c) $\mu=3.63$ and
${\cal T}=256$.
\m
\n{\bf Fig. 2} : Three dimensional views of adiabatic compressions
of a CML with initial length $L=85$, wavenumber
$q=10\left(\lambda=8.5\right)$ and initial state :
$x^0_i=0.45+0.15\sin\left(2\pi q i/L\right).\
\varepsilon=0.667;\ \mu=3.63$. a) Among 120 successive plots
(period ${\cal T}=256$) two destabilizations occur for $L=82$ and
$L=57$. b) The same as the begining of a), the destabilization
observed for $L=82$ is shown in much more details thanks to the
lower plotting period ${\cal T}=32$.
\m
\n{\bf Fig. 3} : A continuous plot of the function $\alpha \left(q,\
\varepsilon\right)$ defined in (2.11) versus $q$ for $\varepsilon
=0.667$ and
$L=100$.
\m
\n{\bf Fig. 4} : A continuous plot of the function $\alpha \left(q,\
\left\{\rho\right\}\right)$ defined in (2.18) versus $q$ for
$L=100$ and linearly decreasing weights $\rho$ of range 5.
\m
\n{\bf Fig. 5} : Stability domain (grey area) of the fixed point of
map $F$ in parameter space $\varepsilon\times\mu$. ($\varepsilon$ :
horizontal axis; $\mu$ : vertical axis).
\m
\n{\bf Fig. 6} : Stability of period two cycles versus
$\varepsilon$ with $\lambda=6$ and $\mu=3.8$. a) Plot of the largest
value of the modulus of the eigenvalues of $Q_2$ given in (6.5)
versus $\varepsilon$. b) Bifurcation diagram of a CML of length
$L=60$ (Horizontal axis : $\varepsilon$) initialized for each
$\varepsilon$ as : $x_i^{\circ}=B+D\cos\left(2\pi i/L\right)$ where
$B$ and $D$ are taken in the vicinity of period two cycle (6.3) of
the reduced map
$F$. For each $\varepsilon$ a transient of $t=200$ has been died and 100
iterations of the CML are drawn with a plotting period ${\cal T}=2$. c)
Bifurcation diagram of reduced  map $F$. For each $\varepsilon$, $Y$
coordinate of $F$ is plotted for 200 iterations after a transient of 200 has
been died.
\eject
\n{\bf Fig. 7} : Plot of the largest value of the modulus of
eigenvalues of $Q_2$ versus $\lambda$ with $\mu=3.8$ and
$\varepsilon=0.667$.
\m
\n{\bf Fig. 8} : A sequence of transitions between period two
states of a CML submitted to adiabatic compressions, with
$\varepsilon=0.667,\ \mu=3.8$ and a stroboscopic period ${\cal
T}=2$. a) Initial state of a CML of length $L=63$ and wavenumber
$q=9\left(\lambda=7\right)$. b) After successive compressions
$L=55$. The lattice is still stable but a modulation of spatial
wavelength may observed. c) Now $L=54$, a superposition of about 20
successive states shows how the destabilization propagates. d) With
$L=54$, after a transient of about 4000 iterations (from c)) the CML
relaxes to a new stable state, strongly modulated in space. e) A
superposition of d) at time $t$ and the state obtained at time
$t+2$, which shows, in fact, a period $T=4$. f), g) A state with
period $T=2$ and wavenumber $q=6$ is restored for $L=48$.
\m
\n{\bf Fig. 9} : Stability domains of $F,\ \left(T=2\right)$ for
various values of $\lambda\left(WL=2,4,6,8\right)$ in parameter
space
$\varepsilon\times\mu$ ($\varepsilon$ : vertical axis, $\mu$ :
horizontal axis).
\m
\n{\bf Fig. 10} : Stability domain of $F,\ \left(T=2\right)$ for
$\mu=3.8$ in parameter space $\varepsilon\times\lambda$
($\varepsilon$ : horizontal axis,
$\lambda$ : vertical axis).
\m
\n{\bf Fig. 11} : With $\mu=3.63,\ \lambda=8$ and $T=4$ ; a) Plot
of the largest value of the modulus of eigenvalues of $Q_4$ versus
$\varepsilon$. b) Bifurcation diagram of a CML of length $L=64$. For
each $\varepsilon$ 50 iterations are plotted after a transient of
$t=200$ has been died.
\m
\n{\bf Fig. 12} : The same as Fig. 11-a but expressed as a function
of spatial wavelength $\lambda$ with $\varepsilon=0.667$ and the
same value of $\mu$.
\m
\n{\bf Fig. 13} : The same scenario as for Fig. 8.a)-g) with
$\varepsilon=0.667,\ \mu=3.63$ and a stroboscopic period ${\cal T}
=4$. a) Initial state of a CML of length $L=72\left(q=8\right)$.b)
The lattice just before destabilization, $L=66$. c) Superposition of
about 20 successive states when $L=65$. d) With $L=65$, the periodic
state obtained after relaxation. e) The state of d) at time $t$ plus
the state at time $t+4$. f) A spatiotemporal periodic state is
restored when $L=52.\ \left(q=5\right)$
\eject
\n{\bf Fig. 14} : Maximal Lyapunov exponants computed along orbits
close, at initial time, the cycle with $T=2,\ \lambda=6,\ \mu=3.8$,
versus
$\varepsilon$ : a) For reduced map $F$ ; for each $\varepsilon$ the
orbit length is 300 after a transient of $t=100$ has been died. b)
For map ${\cal F}$ with 24 Fourier modes ; for each $\varepsilon$
the orbit length is 300 after a transient of $t=50$ has been died.

 \end